\documentclass[10pt,conference]{IEEEtran}

\usepackage[dvipsnames]{xcolor}
\usepackage{tikz}
\usepackage{amsmath}
\usepackage{algpseudocode}
\usepackage[english]{babel}
\usepackage[utf8]{inputenc}
\usepackage{hyphenat}
\usepackage{algorithm}
\usepackage{algpseudocode}
\usepackage{enumitem}
\usepackage{xcolor}
\usepackage{minted}
\usepackage{soul}
\usepackage{balance}
\usepackage{multicol}
\usepackage{amssymb}
\usepackage{bbding}
\usepackage{svg}
\usepackage{ifthen}
\usepackage{csquotes}
\usepackage{caption}
\usepackage{float}
\usepackage[T1]{fontenc}
\usepackage{mathrsfs}
\usepackage{pifont}
\usepackage{booktabs}
\usepackage{array}
\usepackage{stfloats}

\usepackage[colorlinks=true,linkcolor=JungleGreen, citecolor=JungleGreen, anchorcolor=JungleGreen, linkcolor=JungleGreen]{hyperref}

\hypersetup{urlcolor=JungleGreen}

\definecolor{commentcolor}{RGB}{63, 127, 95}
\definecolor{functioncolor}{RGB}{0, 0, 128}
\definecolor{keywordcolor}{RGB}{128, 0, 0}
\definecolor{builtin}{RGB}{51, 51, 255}
\newminted{yaml}{frame=lines,framerule=2pt}
\newboolean{includeappendix}
\setboolean{includeappendix}{true} 

\begin{document}

\title{\Large \bf \textsc{Deptex}: Organization-First \\Open Source Dependency Risk Monitoring}

\author{
\IEEEauthorblockN{Henry Ruckman-Utting, Vrushal Nedungadi, Taiga Okuma, LeTian Wang,}
\IEEEauthorblockN{Stephen Ehebald, Mohammad A. Tayebi}
\IEEEauthorblockA{\textit{School of Computing Science} \\
\textit{Simon Fraser University}\\
Burnaby, Canada \\
\{hjr1, vna28, toa12, lwa118, sea44, tayebi\}@sfu.ca}
}

\maketitle

\begingroup
\renewcommand{\thefootnote}{\*}
\endgroup

\begin{abstract}
Open-source software (OSS) dependencies introduce systemic risks that are difficult to manage at scale. Existing Software Composition Analysis (SCA) and reachability tools generate severe alert fatigue by treating risk as an intrinsic component property, ignoring semantic context and forcing enterprises into rigid compliance frameworks. We present \textsc{Deptex}, an organization-first, graph-based platform treating supply chain risk as emergent. \textsc{Deptex} introduces Execution Path Dominance (EPD), fusing Code Property Graph (CPG) slicing with Large Language Model (LLM) semantic verification to calculate a vulnerability's true operational blast radius. To handle bespoke compliance, \textsc{Deptex} abstracts governance into a programmable ``As Code'' engine, enabling security teams to natively enforce dynamic pull request policies, custom asset tiers, and external API integrations. By shifting from reactive scanning to context-aware governance, \textsc{Deptex} enables proactive, efficient, and aligned supply chain risk management.
\end{abstract}


\section{Introduction} 

Open Source Software (OSS) is foundational to modern development, comprising the majority of shipped code in enterprise environments \cite{synopsys2024}. While leveraging open-source packages accelerates delivery, it introduces systemic risks: vulnerabilities, malicious compromises, and complex license violations. The industry has responded with Software Composition Analysis (SCA) tools and advanced reachability analyzers \cite{snyk2023, endor2024}. These legacy tools successfully provide component-level signals, such as base severity and binary reachability, to identify flawed dependencies.

However, these signals critically lack organizational context. A vulnerability reachable via an unauthenticated, public-facing API receives the exact same severity as one in an offline background script. Consequently, engineering teams suffer from severe Type-B alert fatigue \cite{amreen2019, yoon2021}, overwhelmed by technically reachable but operationally benign alerts. 

Furthermore, real-world governance is rarely as straightforward as globally allowing or denying specific licenses. Enterprises operate under bespoke constraints—such as requiring proprietary legal approvals for packages touching payment gateways. Because legacy platforms rely on rigid checkboxes, organizations expend significant resources building custom internal "wrapper" tools just to filter noise and map alerts to their actual structure.

To address this disconnect and the systemic failures of legacy scanners—namely organizational blind spots, reactive awareness, and stakeholder disconnect—we introduce \textsc{Deptex}, an organization-first software supply chain platform. \textsc{Deptex} posits that supply chain risk is \emph{emergent}—arising from how dependencies, assets, and operational environments interact.

This paper presents the design and architecture of \textsc{Deptex}. We make the following core contributions, which concurrently serve as our primary architectural design goals:
\begin{itemize}[leftmargin=*, label=$\triangleright$]
    \item \textbf{Programmable Governance (Security ``As Code''):} We abstract compliance into an extensible, declarative model, allowing organizations to integrate bespoke external enterprise APIs, define highly specific dependency constraints, and natively enforce dynamic pull request gatekeeping.
    \item \textbf{Context-Aware Prioritization (Depscore):} We introduce Execution Path Dominance (EPD), an algorithm fusing deterministic Framework-Forward Reachability with Large Language Model (LLM) semantic verification to calculate the true operational blast radius of a vulnerability, drastically reducing false-positive urgency.
\end{itemize}

\section{Background and Problem Context}

\subsection{Related Work and Industry Limitations}
\textbf{Alert Fatigue and Binary Reachability.} The explosive growth of the software supply chain necessitates automated SCA tools \cite{synopsys2024}. However, extensive research indicates these tools are severely hampered by notification fatigue \cite{amreen2019, alfadel2021, yoon2021}. Developers frequently ignore remediation requests due to high volume, with only $\sim$32\% of automated pull requests being merged \cite{mirhosseini2017}. While vulnerability detection is mature, effective prioritization remains unsolved \cite{alsharif2023}. 

Modern commercial tools \cite{endor2024} introduced Advanced Reachability Analysis using Call Graphs and Abstract Syntax Trees (AST). Yet, these tools still yield binary outputs (True/False) lacking semantic environmental context. Treating an offline script vulnerability with the same urgency as a public API failure fundamentally fails to reduce Type-B alert fatigue.

\vspace{1mm}
\noindent \textbf{The Operational Burden of Governance.} Because current tools fail to semantically contextualize reachability, the burden of filtering noise falls entirely on the organization. Enterprises must build brittle, bespoke internal tooling to wrap around legacy scanners to map alerts to their actual infrastructure. This paradigm is highly inefficient, decentralizes the security posture, and fails to scale with complex compliance requirements.

\subsection{Problem Formulation}
Despite adopting state-of-the-art tools, organizations remain reactive. Only 15\% of CISOs claim full visibility into their open-source usage \cite{PanoraysWhitePaper}, and 72\% of professionals identify supply chain security as a critical blind spot \cite{SecurityBoulevard}. We identify three core problems facing enterprise supply chain security:

\begin{itemize}[leftmargin=*, label=$\triangleright$]
  \item \textbf{Organizational Blind Spots:} Open-source usage is fragmented across repositories, leaving no single stakeholder with a unified view of organizational risk.
  \item \textbf{Reactive Awareness:} Risks generally become visible only after a scan completes or an external disclosure occurs, rather than being proactively intercepted during the continuous integration (PR) process.
  \item \textbf{Stakeholder Disconnect:} Legacy signals are directed as raw alerts to developers. This isolates engineering managers and compliance officers, resulting in poorly coordinated decision-making.
\end{itemize}

\begin{figure*}[b]
  \centering
  \includegraphics[width=\textwidth]{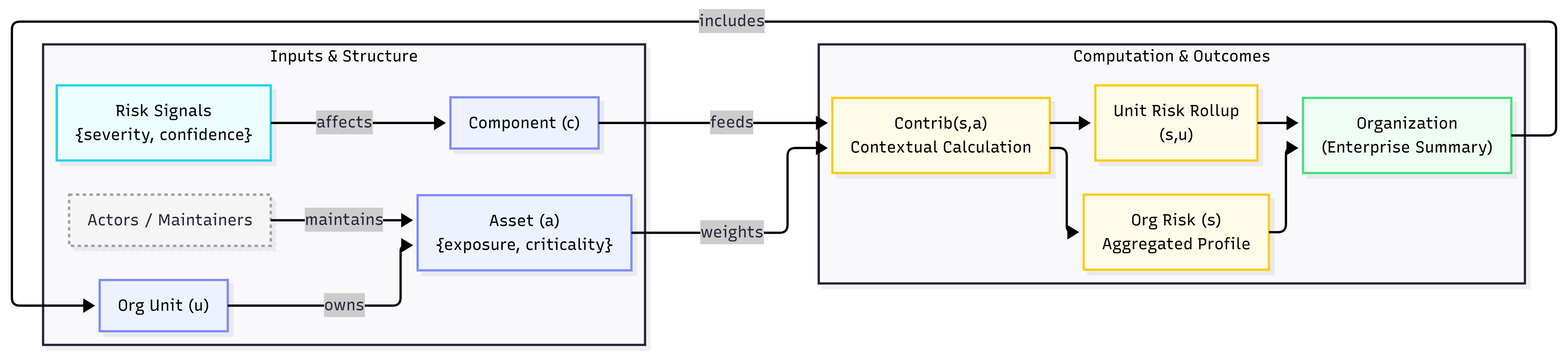}
  \caption{The hierarchical risk propagation model. Vulnerability signals ($\mathsf{Signal}$) are quantified via contextual contributions ($\mathsf{Contrib}(s,a)$) and aggregated upwards through the organizational structure.}
  \label{fig:conceptual_model}
\end{figure*}

\section{The Conceptual Framework}
\label{sec:conceptual-model}

We posit that supply chain risk is \emph{emergent}: it arises from how dependencies, assets, owners, and operational contexts interact. To operationalize this, our conceptual model treats the organization as the primary unit of analysis, supporting reasoning about risk propagation and governance metrics.

\subsection{Typed Property Graph Abstraction}
We model supply chain information as a typed property graph $G = (V, E, \tau_V, \tau_E, \mathcal{A})$, where $V$ is the set of nodes, $E \subseteq V \times V$ is the set of directed edges, $\tau_V$ and $\tau_E$ assign types to nodes and edges, and $\mathcal{A}$ defines attributes.

The core node types ($\mathcal{T}_V$) map the enterprise ecosystem: \textbf{Organizations} ($\mathsf{Org}$) and \textbf{Units} ($\mathsf{Unit}$) define the hierarchy; \textbf{Assets} ($\mathsf{Asset}$) represent deployables; \textbf{Components} ($\mathsf{Comp}$) represent dependencies; \textbf{Actors} ($\mathsf{Actor}$) denote maintainers; and \textbf{Risk Signals} ($\mathsf{Signal}$) represent observed vulnerabilities.

Edges ($\mathcal{T}_E$) map operational realities: $\mathsf{Org} \xrightarrow{\mathsf{contains}} \mathsf{Unit}$, $\mathsf{Unit} \xrightarrow{\mathsf{owns}} \mathsf{Asset}$, and $\mathsf{Asset} \xrightarrow{\mathsf{depends\_on}} \mathsf{Comp}$ (encoding scope and directness attributes). Vulnerabilities map as $\mathsf{Signal} \xrightarrow{\mathsf{affects}} \mathsf{Comp}$.

\subsection{Integrity Constraints and Impact}
The model enforces structural constraints to support consistent reasoning, such as \textit{Ownership completeness}. If an asset lacks an owner, it exhibits a governance risk gap. For asset $a$:
\[
\begin{aligned}
\mathsf{OwnershipGap}(a) \triangleq\;& \neg \exists u \in V:\; \tau_V(u)=\mathsf{Unit} \\
& \wedge\; (u,a)\in E \;\wedge\; \tau_E(u,a)=\mathsf{owns}
\end{aligned}
\]

Given a risk signal $s$, we define the topological blast radius by deriving the affected assets and units via dependency reachability:
\[
\begin{aligned}
\mathsf{AffectedAssets}(s) \triangleq\;
\left\{ a \in V \;\middle|\; \tau_V(a)=\mathsf{Asset} \right. \\
\left. \wedge\ \exists c \in V:\; \tau_V(c)=\mathsf{Comp} \wedge (s,c)\in E \wedge (a,c)\in E \right\}
\end{aligned}
\]
\[
\begin{aligned}
\mathsf{AffectedUnits}(s) \triangleq\;
\left\{ u \in V \;\middle|\; \tau_V(u)=\mathsf{Unit} \right. \\
\left. \wedge\ \exists a \in \mathsf{AffectedAssets}(s):\; (u,a)\in E \right\}
\end{aligned}
\]

\subsection{Risk Propagation and Governance}
A central design principle is that a component-level signal does not directly yield actionable organizational risk. Risk emerges via contextualization. For a signal $s$ and an affected asset $a$, we define a contextual contribution using a monotone function $\phi$:
\[
\begin{aligned}
\mathsf{Contrib}(s,a) \triangleq\;
\phi\big(\mathsf{sev}(s),\mathsf{conf}(s), \mathsf{direct}(a,s), \\
\mathsf{scope}(a,s), \mathsf{exposure}(a),\mathsf{critical}(a), \\
\mathsf{OwnershipGap}(a)\big)
\end{aligned}
\]

This risk is then mathematically aggregated upward to the team ($\mathsf{Unit}$) and enterprise ($\mathsf{Org}$) levels to provide a prioritized, organization-first ranking:
\[
\begin{aligned}
\mathsf{UnitRisk}(s,u) \triangleq\;
\mathrm{Agg}\big(\{ \mathsf{Contrib}(s,a) \mid (u,a)\in E, \\
\tau_E(u,a)=\mathsf{owns} \}\big)
\end{aligned}
\]
\[
\mathsf{OrgRisk}(s) \triangleq
\mathrm{Agg}\left(\{\mathsf{UnitRisk}(s,u)\;|\;\tau_V(u)=\mathsf{Unit}\}\right)
\]

By abstracting risk in this manner, the graph natively supports actionable governance metrics, including organizational exposure ($|\mathsf{AffectedAssets}(s)|$), cross-team coordination overhead ($|\mathsf{AffectedUnits}(s)|$), and specific ownership gaps within the blast radius.

\section{System Architecture}
\label{sec:architecture}

The \textsc{Deptex} platform operationalizes the conceptual Typed Property Graph through a suite of automated, context-aware features. Unlike legacy tools that treat repositories as isolated silos, \textsc{Deptex} is fundamentally designed around the organizational hierarchy. Every architectural decision—from the ingestion of SBOMs to the routing of notifications—is mapped against the graph's entity relationships. By defining which $\mathsf{Unit}$ owns which $\mathsf{Asset}$ and assigning business-logic modifiers, the platform ensures that risk calculations are judged by their operational blast radius within the enterprise structure.

\begin{figure}[b]
  \centering
  \vspace{-1em} 
  \includegraphics[width=1\linewidth]{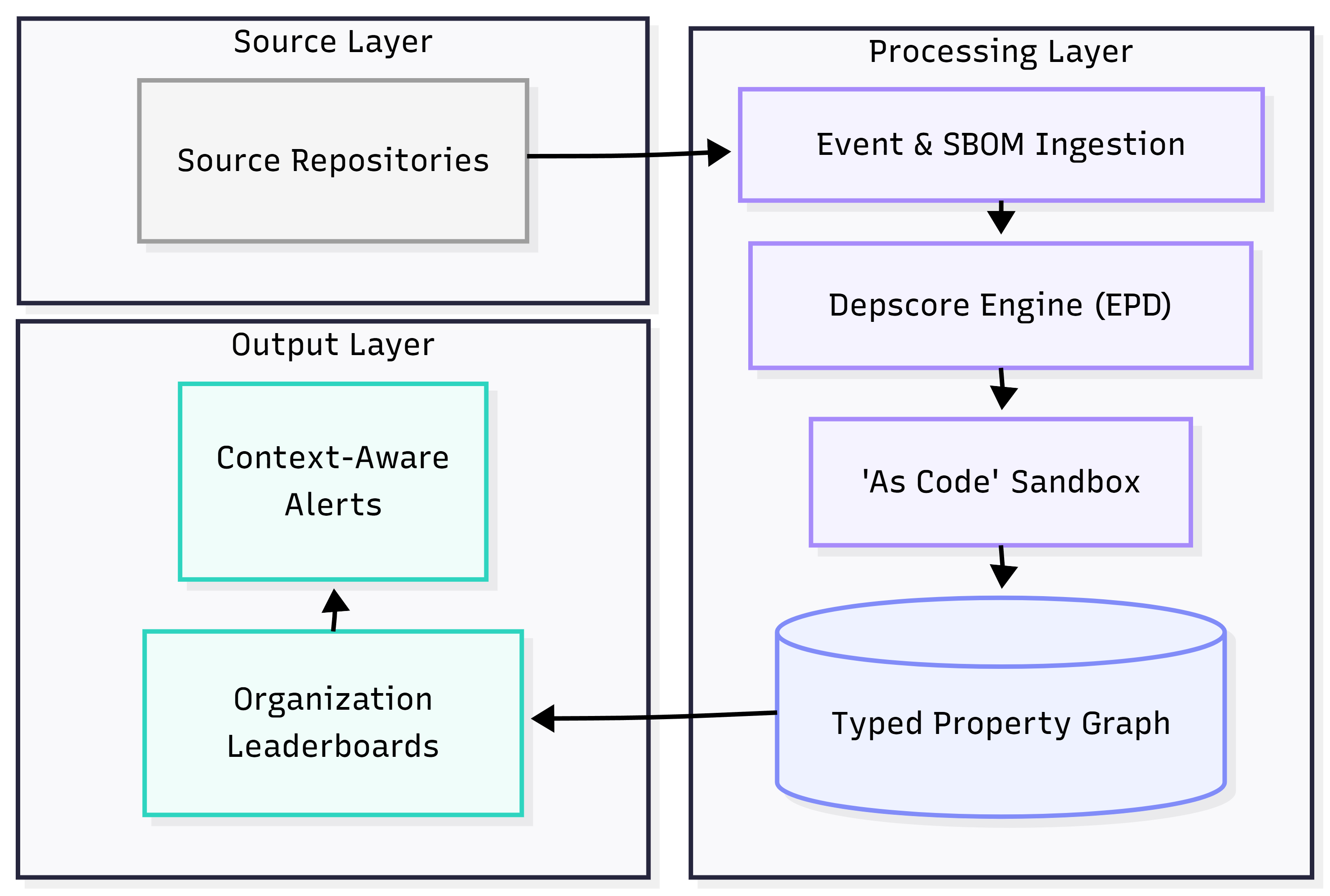}
  \vspace{-1em} 
  \caption{System architecture.}
  \vspace{-1.5em} 
  \label{fig:architecture}
\end{figure}

\subsection{Pillar I: Programmable Governance}
Legacy Application Security Posture Management (ASPM) tools typically utilize rigid RBAC and binary compliance states (Compliant vs. Non-Compliant). \textsc{Deptex} reverses this paradigm with a Security ``As Code'' execution engine—an AI-assisted JavaScript sandbox that runs natively within the platform to evaluate organizational data against programmable logic.

To faithfully represent the $\mathsf{Org}$, $\mathsf{Unit}$, and $\mathsf{Actor}$ entities, \textsc{Deptex} introduces granular modeling:
\begin{itemize}[leftmargin=*, label=$\triangleright$]
    \item \textbf{Asset Tiers:} Projects are categorized into custom tiers (e.g., \textit{Payment Gateway}, \textit{Public API}). Each tier carries an \textit{Importance Modifier}, a scalar value directly influencing the $\mathsf{Contrib}(s,a)$ risk calculation.
    \item \textbf{Custom Compliance Statuses:} Users define a bespoke taxonomy of states (e.g., \textit{Legal Hold}, \textit{Quarantined}) aggregated via an interactive UI, allowing leadership to instantly assess the distribution of $\mathsf{UnitRisk}$.
\end{itemize}

The cornerstone of this pillar is the execution engine, supporting four primary contexts: (i) \textbf{Status as Code}, programmatically triggering compliance transitions based on asset attributes; (ii) \textbf{Policy as Code}, enforcing standards on individual $\mathsf{Comp}$ metadata; (iii) \textbf{Pull Request (PR) as Code}, acting as a CI/CD gatekeeper by evaluating the delta of a proposed change against Asset Tiers; and (iv) \textbf{Notifications as Code}, dispatching event-driven alerts (e.g., PagerDuty, Slack) only after verifying structural reachability and criticality.

\subsection{Pillar II: Context-Aware Prioritization (Depscore)}
To operationalize the $\mathsf{Contrib}(s,a)$ function, \textsc{Deptex} introduces \textbf{Execution Path Dominance (EPD)}, or the Depscore. The engine abandons static severity in favor of a three-phase calculation:

\begin{figure}[t]
  \centering
  \includegraphics[width=\linewidth]{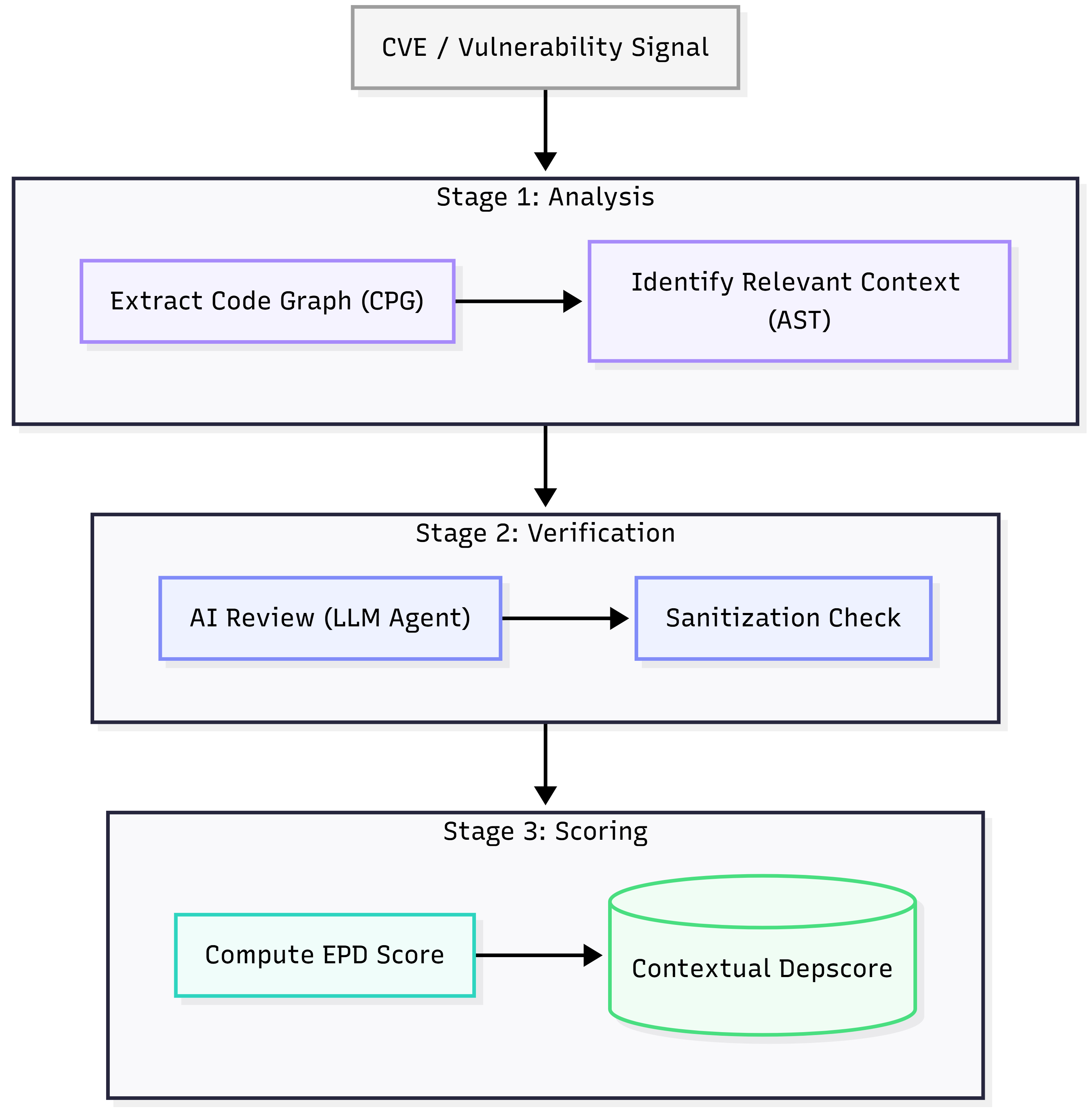}
  \caption{The Depscore calculation pipeline.}
  \vspace{-0.5em} 
  \label{fig:depscore_pipeline}
\end{figure}

\textbf{1) Structural Slicing:} Using the \texttt{atom} parser, the system generates Code Property Graph (CPG) slices. This deterministic phase extracts the exact data-flow path connecting the $\mathsf{Asset}$ entry point to the vulnerable sink in the target $\mathsf{Comp}$, calculating the path depth ($d$)—the number of execution hops in the chain.

\textbf{2) Semantic Verification:} To prevent context bloat, the system extracts only the specific functional code blocks identified in the slice. These are passed to a constrained LLM agent utilizing strictly typed JSON schemas. The agent classifies the entry point exposure ($W_{entry}$), weighting public-facing APIs (1.0) higher than internal background tasks (0.1). Simultaneously, it performs semantic reasoning to verify if custom sanitization logic ($\mathsf{is\_sanitized}$) neutralizes the vulnerability's precondition.

\textbf{3) Geometric Decay Calculation:} The final Depscore models ``path friction.'' As malicious payloads cross function boundaries, exploitation probability decreases. We model this as:
\[ EPD = W_{entry} \times \alpha^d \]
where $\alpha$ is the attenuation factor (e.g., 0.85). If the agent concludes $\mathsf{is\_sanitized}$ is true, the $EPD$ is strictly forced to $0.0$, mathematically eliminating the false positive from the $\mathsf{Unit}$ leaderboard.

\section{Evaluation and Comparative Analysis}
\label{sec:evaluation}
We evaluate \textsc{Deptex} through operational scenarios demonstrating its programmable governance, followed by a feature-level comparison against Dependabot, Dependency-Track, and Snyk.

\subsection{Operational Use Cases and Validation}
We present four operational scenarios showing how programmable governance and contextual prioritization resolve the systemic failures of legacy SCA tools.

\subsubsection{Use Case 1: External API PR Gatekeeping}
\textbf{Scenario:} A heavily regulated enterprise requires proprietary legal review before any new open-source library is merged into production, forcing developers to manually open legal tickets and creating severe CI/CD bottlenecks.

\noindent \textbf{Outcome (\textsc{Deptex} Solution):} Using the \textbf{PR as Code} execution context, a DevSecOps architect authors a policy that intercepts dependency graph additions, packages the library's metadata, and queries the enterprise's internal Legal API. If the API returns an ``Unapproved'' status, the PR is dynamically blocked with an explanatory comment.

\subsubsection{Use Case 2: Context-Aware Alert Routing}
\textbf{Scenario:} A zero-day vulnerability simultaneously affects a \textit{Payment Gateway} (Asset Tier 1) and an \textit{Internal Analytics Dashboard} (Asset Tier 3), causing standard reachability tools to page all maintainers across both projects and generate Type-B alert fatigue for the internal tools team.

\noindent \textbf{Outcome (\textsc{Deptex} Solution):} Using \textbf{Notifications as Code}, the alert routing script evaluates Asset Tier before dispatching webhooks. The Tier 1 asset immediately triggers a PagerDuty incident for the payments team; the Tier 3 asset suppresses the page and silently opens a Jira backlog ticket instead.

\subsubsection{Use Case 3: Prioritization via Depscore (EPD)}
\textbf{Scenario:} A critical CVSS 9.8 vulnerability in a ubiquitous parsing library flags 10 reachable repositories via a commercial binary reachability scanner.

\noindent \textbf{Outcome (\textsc{Deptex} Solution):} The Depscore engine analyzes the 10 structurally reachable assets. In 8 repositories, the semantic LLM agent identifies that the vulnerable package is invoked exclusively via an offline background batch script (low $W_{entry}$) and sits 6 function hops deep ($d=6$), yielding a negligible EPD that safely downgrades the alerts. The remaining 2 repositories show direct, unauthenticated public API exposure and receive an EPD of 92.

\subsection{Tool Comparison}
To contextualize \textsc{Deptex} against the broader ecosystem, we compare its capabilities with established industry standards: Dependabot (automated remediation), Dependency-Track (SBOM-centric inventory), and Snyk (commercial SaaS scanning). As summarized in Table \ref{tab:comparison}, while baseline features are well-supported across modern tools, \textsc{Deptex} addresses a persistent gap in \emph{organization-level} contextual risk governance.
\begin{table}[h]
\centering
\footnotesize
\caption{Feature Comparison of SCA Tools}
\label{tab:comparison}
\renewcommand{\arraystretch}{1.2}
\begin{tabular}{|l|c|c|c|c|}
\hline
\textbf{Feature} & \textbf{Depend.} & \textbf{D-Track} & \textbf{Snyk} & \textbf{Deptex} \\ \hline
CI/CD \& Ticketing Integrations & \ding{51} & \ding{51} & \ding{51} & \ding{51} \\
License \& Compliance Auditing & \ding{55} & \ding{51} & \ding{51} & \ding{51} \\
Org-Wide Portfolio View & \ding{55} & \ding{51} & \ding{51} & \ding{51} \\
Self-Hosted Deployment & Partial & \ding{51} & \ding{55} & \ding{51} \\
Structural Reachability (CPG) & \ding{55} & \ding{55} & Partial & \ding{51} \\
Contextual Vulnerability Scoring & \ding{55} & \ding{55} & \ding{55} & \ding{51} \\
Programmable ``As Code'' Policy & \ding{55} & \ding{55} & \ding{55} & \ding{51} \\ \hline
\end{tabular}
\end{table}
While Dependabot provides seamless baseline PR automation, it fragments situational awareness by generating hundreds of isolated alerts across independent repositories. Dependency-Track offers excellent self-hosted SBOM aggregation and license auditing but relies entirely on static, non-contextual severity metrics (e.g., base CVSS). Snyk provides advanced scanning and partial reachability insights but operates as a SaaS platform with rigid, vendor-defined compliance frameworks that struggle to accommodate bespoke enterprise logic.

\textsc{Deptex} achieves parity on baseline features but differentiates itself in how risk is measured and enforced: by calculating risk contextually against an asset's operational exposure, it eliminates the false-positive noise inherent to Dependency-Track. Its self-hosted architecture further ensures that sensitive dependency data remains internal, allowing the ``As Code'' governance engine to securely query proprietary enterprise APIs and enforce bespoke workflows that commercial SaaS solutions cannot support.

\section{Conclusion and Future Work}
\label{sec:conclusion}

We presented \textsc{Deptex}, an organization-first software supply chain platform. By treating risk as an emergent property of the organizational graph, \textsc{Deptex} addresses the systemic failures of legacy SCA tools. Execution Path Dominance (EPD) provides contextual prioritization, while our Security ``As Code'' engine allows enterprises to natively enforce bespoke compliance. Ultimately, \textsc{Deptex} shifts security from reactive, component-level scanning to proactive, organizationally aligned governance.

Future work will focus on longitudinal enterprise case studies to quantify the reduction in Type-B alert fatigue. Additionally, integrating Static Application Security Testing (SAST) and secrets detection into the software will evolve \textsc{Deptex} into a comprehensive Application Security Posture Management (ASPM) platform.

\vspace{1mm}
\noindent\textbf{Tool Availability:} \textsc{Deptex} is available as a web application at \url{https://deptex.dev}. To support reproducibility and future research, the complete platform is open source at \url{https://github.com/deptex/deptex}.

\newpage

\medskip
\onecolumn \begin{multicols}{2}

\end{multicols}
\end{document}